\documentclass[a4paper]{article}

\usepackage{INTERSPEECH2020}
\usepackage{xcolor}
\usepackage{booktabs}

\usepackage{array,multirow,graphicx}
\usepackage{float}
\usepackage{enumitem} 
\usepackage{hyperref}

\title{Cross-Domain Adaptation of Spoken Language Identification for Related Languages: The Curious Case of Slavic Languages}

\name{Badr M. Abdullah$^{1, 2}$, Tania Avgustinova$^{2}$, Bernd M\"obius$^{2}$, Dietrich Klakow$^{1, 2}$}
\address{
  $^1$Spoken Language Systems (LSV) \hspace{1cm} $^2$Department of Language Science and Technology\\
   Saarland Informatics Campus, Saarland University, 66123 Saarbrücken, Germany
  }
\email{\{babdullah|dietrich\}@lsv.uni-saarland.de, \{avgustinova|moebius\}@lst.uni-saarland.de}

\begin{document}

\maketitle
\begin{abstract}
State-of-the-art spoken language identification (LID) systems, which are based on end-to-end deep neural networks, have shown remarkable success not only in discriminating between distant languages but also between closely-related languages or even different spoken varieties of the same language. However, it is still unclear to what extent neural LID models generalize to speech samples with different acoustic conditions due to domain shift. In this paper, we present a set of experiments to investigate the impact of domain mismatch on the performance of neural LID systems for a subset of six Slavic languages across two domains (read speech and radio broadcast) and examine two low-level signal descriptors (spectral and cepstral features) for this task. Our experiments show that (1) out-of-domain speech samples severely hinder the performance of neural LID models, and (2) while both spectral and cepstral features show comparable performance within-domain, spectral features show more robustness under domain mismatch. Moreover, we apply unsupervised domain adaptation to minimize the discrepancy between the two domains in our study. We achieve relative accuracy improvements that range from 9\% to 77\% depending on the diversity of acoustic conditions in the source domain.

\end{abstract}
\noindent\textbf{Index Terms}: spoken language identification, Slavic languages, deep neural networks, unsupervised domain adaptation

\section{Introduction}

Spoken language identification, henceforth LID, is the problem of determining the identity of the language in a spoken utterance \cite{li2013spoken}. In today's globalized world, LID systems can facilitate a wide range of cross-lingual speech and communication technologies such as spoken language translation \cite{waibel2000multilinguality, fugen2007simultaneous, bangalore2012real} and multilingual spoken document retrieval \cite{chelba2008retrieval}. Furthermore, LID-aware transfer of language resources has been shown to be effective for multilingual ASR in low-resource settings \cite{muller2016language, muller2015using, nguyen2014multilingual, cutler2014language}.


Earlier work has addressed the LID task using the so-called phonotactic approach. In this paradigm, the acoustic signal is first transduced into a sequence of discrete symbols (e.g., phones), then probabilistic models are utilized to obtain language likelihoods \cite{lamel1994language, li2005phonotactic}. This approach has been outperformed by acoustic approaches that are based on Gaussian Mixture Models (GMMs) and the i-vector framework which has been applied to speaker and language identification \cite{kenny2010bayesian, garcia2011analysis, martinez2011language, su2016factor}. Currently, end-to-end deep neural networks (DNNs) are predominant for LID and outperform GMMs, especially for short utterances \cite{mateju2018using, shen2018feature, shon2018convolutional, lopez2014automatic, gonzalez2014automatic}. 

The findings of the popular language guessing game, the Great Language Game \cite{skirgaard2017some}, have shown that discriminating between closely-related languages is a difficult task for humans. On the other hand, neural LID models have shown striking performance discriminating between spoken varieties of Arabic  \cite{bulut2017utd, gelly2016language, shon2018convolutional}, Slavic languages \cite{mateju2018using}, and languages in accented speech samples from multilingual speakers \cite{titus2020improving}. For instance, the best neural LID model in \cite{mateju2018using} has reported an error rate as low as 1.2\% when discriminating between 11 Slavic languages. Generally speaking, the impressive performance of DNN-based LID reported in the literature gives the impression that LID is almost a solved problem.
 

However, previous works have developed their models using disjoint splits of the same dataset where the training and evaluation samples have similar, if not identical, acoustic conditions (i.e., same domain). The impact of \textit{dataset-bias} \cite{tzeng2017adversarial} on LID robustness has not yet been investigated with a systematic evaluation across datasets. In this paper we aim to fill this gap and focus on the challenging case of LID for short utterances of related languages (i.e., Slavic languages) in a cross-domain setting.  We investigate the following questions:

\begin{itemize}
    \item \textbf{RQ1} \hspace{0.1cm} To what degree do neural LID models for related languages generalize to another domain with different acoustic conditions?
    \item  \textbf{RQ2} \hspace{0.1cm} Are different low-level speech features equally robust under domain mismatch?
    \item \textbf{RQ3} \hspace{0.1cm} Can we adapt LID models to a new domain without using labelled data in the new domain? If yes, what are the factors that affect the adaptability of the model?
\end{itemize}


\noindent
To address these research questions, we conduct a series of LID experiments with datasets from two domains: (1) Read speech recordings from the Slavic subset of the GlobalPhone speech database \cite{schultz2013globalphone}, and (2) Slavic broadcast recordings collected and distributed in \cite{mateju2018using,nouza2016asr} for LID (\textbf{RQ1}). We also compare the performance of spectral (MFSCs) and cepstral (MFCCs) speech features within- and across-domain (\textbf{RQ2}). Finally, we apply adversarial domain confusion  \cite{ganin2015unsupervised} to adapt our model to a target domain, analyze predictions from the adapted model, and visualize its representations compared to the baseline (\textbf{RQ3}). 

\noindent

\section{LID with Deep Neural Networks}

\subsection{Problem Definition}
We define the LID task as a discriminative sequence classification problem. First, a variable-length utterance is transformed by an acoustic front-end into a sequence of acoustic observations $\mathbf{X} = (\mathbf{x}_1, \dots, \mathbf{x}_T)$, where $\mathbf{x}_t \in \mathbb{R}^k $ is a low-level feature vector  at timestep $t$. Given a  sequence  $\mathbf{X}$, the goal is to predict the spoken language $\hat{y}$. Using a deep neural network as a classification model, the LID problem can be defined as 

\begin{equation} \hat{y} =  \underset{y \in \mathcal{Y} }{\arg\max} \: P(y \; | \; \mathbf{X};\;  \boldsymbol{\theta})\end{equation}
where $\mathcal{Y}$ is a finite set of languages, $\boldsymbol{\theta}$ is the model's parameters learned in a supervised approach, and  $P(y  |  \mathbf{X}; \boldsymbol{\theta})$ represents a posterior probability of the language label $y$.

\subsection{LID Model Overview}
Our LID model consists of a 1D 3-layer convolutional network followed by 2-layer fully-connected feed-forward network as schematized in Fig. \ref{fig:modelscheme}(a). We refer to the convolutional block as a high-level feature extractor $G_f$ that transforms the input sequence $\mathbf{X}$ into a $D$-dimensional feature vector $\mathbf{f} \in  \mathbb{R}^D$, i.e. $\mathbf{f} = G_f(\mathbf{X}; \boldsymbol{\theta}_f)$. Then, the feed-forward layers transform $\mathbf{f}$ into a logit vector $\mathbf{\hat{y} \in \mathbb{R}^{|\mathcal{Y}|}}$ via a series of non-linear transformations, i.e. $\mathbf{\hat{y}} = G_y(\mathbf{f}; \boldsymbol{\theta}_y)$, followed by a softmax function that maps $\mathbf{\hat{y}}$ into a probability distribution over the language space. We refer to the fully-connected block of the model $G_y$ as a language classifier. The parameters of the network $\boldsymbol{\theta}_f$ and $\boldsymbol{\theta}_y$ are learned jointly in an end-to-end approach given a dataset $\mathcal{D}_\mathcal{S} = \{(\mathbf{X}_{i}, y_{i})\}_{i=1}^{N_\mathcal{S}}$ of ${N_\mathcal{S}}$ labelled samples in one domain. The objective function is to minimize
\begin{equation}J(\boldsymbol{\theta}_f, \boldsymbol{\theta}_y) = \sum_{(\mathbf{X}_i, y_i) \in \mathcal{D}_\mathcal{S}}^{}L_y\Big(G_y\big(G_f(\mathbf{X}_i; \boldsymbol{\theta}_f); \boldsymbol{\theta}_y\big), y_i\Big) \end{equation}
where $L_y$ is the loss of the language classifier.

\subsection{Domain Adaptive LID}
In this paper, we explore a well-established domain adaptation technique that has been successfully applied to many vision and speech recognition problems \cite{ganin2015unsupervised, shinohara2016adversarial, meng2017unsupervised}. This technique aims to minimize the discrepancy between two domains given a dataset $\mathcal{D}_\mathcal{T} = \{\mathbf{X}_{i}\}_{i=1}^{N_\mathcal{T}}$  of ${N_\mathcal{T}}$  unlabelled samples in the target domain, in addition to the source labelled samples $\mathcal{D}_\mathcal{S}$.  

\begin{figure}[t]
  \centering
  \includegraphics[width=0.95\linewidth]{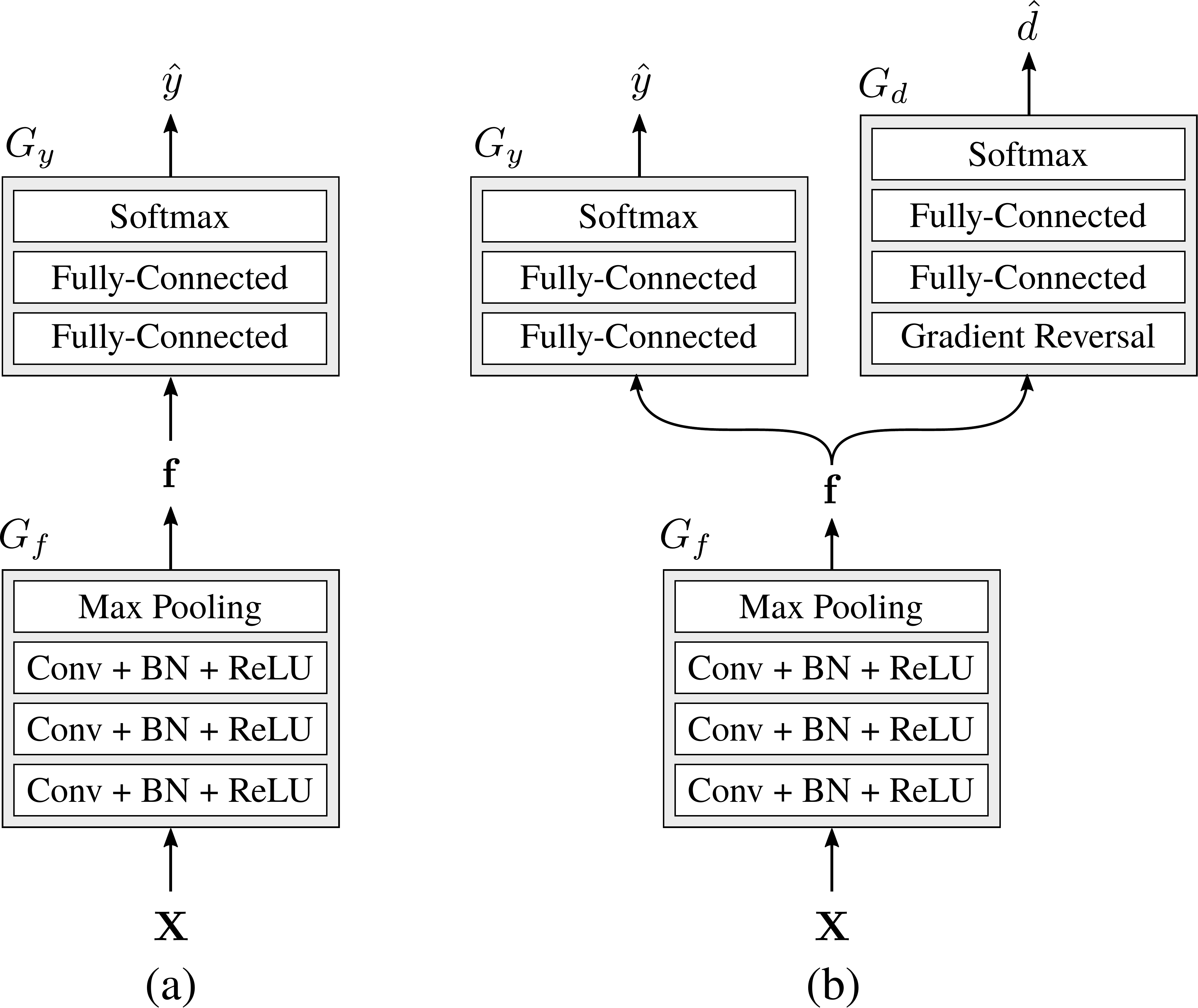}
  \caption{A schematic view of our models: (a) Non-adaptive, and (b) Domain adaptive LID with adversarial classifier $G_d$. }
  \label{fig:modelscheme}
\end{figure}

To improve the LID model's out-of-domain generalization, the feature representations emerging from the model should be both language-discriminative and domain-invariant. This objective can be achieved if the model is encouraged during training to build up representations that are good predictors of the spoken language but do not encode domain-related information. To this end, a  fully-connected feed-forward block $G_d$ is added to the network to predict the domain given $\mathbf{f}$ (see Fig. \ref{fig:modelscheme}(b)). We view $G_d$ as a domain classifier with a separate set of parameters $\boldsymbol{\theta}_d$ which are learned by exploiting the domain labels of source and target samples. That is, each training sample in the source domain $(\mathbf{X}_i, y_i)$ is augmented with a domain label $d_i = 0$, while each training sample in the target domain $\mathbf{X}_j$ is augmented with a domain label $d_j = 1$. We seek the parameters $\boldsymbol{\theta}_d$  that minimize the loss of the domain classifier. On the other hand, the feature extractor $G_f$ is trained such that $\mathbf{f}$ is uninformative for the domain classifier. Thus, we seek the parameters $\boldsymbol{\theta}_f$ that maximize the domain classifier loss. This procedure is an instance of adversarial learning where different blocks in the network are trained with competing objectives. The overall objective function is to minimize 
\begin{multline}
J(\boldsymbol{\theta}_f, \boldsymbol{\theta}_y, \boldsymbol{\theta}_d) =  \sum_{(\mathbf{X}_i, y_i) \in \mathcal{D}_\mathcal{S}}^{}L_y\Big(G_y\big(G_f(\mathbf{X}_i; \boldsymbol{\theta}_f); \boldsymbol{\theta}_y\big), y_i\Big) \\ - \lambda \sum_{(\mathbf{X}_i, d_i) \in (\mathcal{D}_\mathcal{S} \cup \mathcal{D}_\mathcal{T})}^{} L_d\Big(G_d\big(G_f(\mathbf{X}_i; \boldsymbol{\theta}_f); \boldsymbol{\theta}_d\big), d_i\Big)
\end{multline}
where $L_y$ is the loss of the language classifier, $L_d$ is the loss of the domain classifier, and $\lambda$ is a parameter that controls the contribution of the domain classifier's loss to the overall loss. In practice, this adversarial loss is realized with a special layer that reverses the direction of the gradient signal coming from the domain classifier's loss into the feature extractor during backpropagation, which is referred to as a gradient reversal layer.  We refer the reader to the original paper for a detailed overview of the training procedure \cite{ganin2015unsupervised}. 

\section{Experimental Setup}
\subsection{Datasets for Slavic LID}
\textbf{GlobalPhone Read Speech (GRS)} \hspace{0.1cm} We use the Slavic portion of the multilingual GlobalPhone speech database \cite{schultz2013globalphone} which includes  read speech recordings from native speakers of six Slavic languages: Bulgarian (BUL), Croatian (HRV), Czech (CZE), Polish (POL), Russian (RUS), and Ukrainian (UKR). The utterances vary in length and quality across languages. We set the minimum utterance length to 3 seconds and segment longer utterances into non-overlapping 3-second speech segments. Our final training subset consists of 8,000 utterances per language. We use the same splits as in \cite{9053144}. \vspace{0.15cm}

\noindent
\textbf{Radio Broadcast Speech (RBS)} \hspace{0.1cm} A large collection of Slavic recordings were collected by harvesting online radio broadcasts in \cite{mateju2018using, nouza2016asr}. The original dataset contains recordings for 11 Slavic languages. We use the same subset of six languages as in the GRS dataset. The extracted utterances are either segments of professional news reports or of spontaneous speech during discussions. Occasionally, the utterances include background music and different sorts of acoustic noise. We sample 8,000 and 500 utterances per language from the training split as our training and validation sets, respectively. This dataset does not include any speaker IDs. Thus, we cannot confirm whether  training and evaluation speakers are disjoint. 
\subsection{Low-level Feature Extraction}
In our experiments, we use the first 13 coefficients of MFSCs and MFCCs, with the zeroth coefficient being the average frame energy, as low-level speech features. While previous works usually refer to MFSCs as mel-filterbanks \cite{shon2018convolutional}, we use the term MFSCs to refer to mel-frequency spectral features that are correlated \cite{mohameddeep}. Since both datasets in our study are sampled at 16~kHz, we extract frames of 400 samples with 160 samples overlap, which corresponds to 25~ms and 10~ms, respectively. We normalize the features to have utterance-level zero mean and unit variance.

\subsection{Model Architecture and Hyperparameters}
\textbf{CNN Architecture} \hspace{0.1cm} We use 1D 3-layer convolution over the temporal dimension with 128, 256, and 512 filters and widths of 5, 10, and 10 for each layer and keep stride step at 1. We apply batch normalization and ReLU non-linearity following each convolutional operation. We apply max pooling to downsample the representation only at the end of the convolution block. For the language classifier, we use 2 fully-connected layers (512 $\rightarrow$ 512 $\rightarrow$ 6) before the softmax for both the non-adapted and the adapted LID models. \vspace{0.1cm}

\noindent
\textbf{Domain-Adaptive Model} \hspace{0.1cm} For our adapted LID models, we use a 3-layer feed-forward network (512 $\rightarrow$ 1024 $\rightarrow$ 1024 $\rightarrow$ 2) as the domain classifier. For the adaptation factor $\lambda$, we use a gradually increasing value $\in [0, 1]$ to suppress the noise from the feature extractor during the initial phase of the training procedure. We experiment with two variants of the domain-adaptive model: (1) \textsc{DA-LID I}: an identical configuration to \cite{ganin2015unsupervised}, where the convolutional block of the model is considered as the feature extractor, and (2) \textsc{DA-LID II}: we consider the feature extractor as the convolutional block as well as the first layer of the fully-connected block; thus, the reversed gradient signal from the domain classifier is back-propagated into all layers of the network except the final layer before the softmax of the language classifier. \vspace{0.1cm}

\noindent
\textbf{Training Details} \hspace{0.1cm} We use cross-entropy loss for both $L_y$ and $L_d$. The ADAM optimizer is used with learning rate of 0.001. We train our models with a batch size of 256 for 50 epochs and observe the validation performance during training.

\vspace{0.1cm} 

\noindent
\textbf{Implementation} \hspace{0.1cm} We use PyTorch to implement the LID models and make our code  publicly available.\footnote{\url{https://github.com/uds-lsv/da-lang-id }} 

\section{Experimental Results}
We now present and discuss the results of our experiments. To make the results comparable across datasets and prevent undesirable effects due to utterance length mismatch, we train and evaluate each of our LID models on 3-second utterances. Since the GRS evaluation data is imbalanced, we use balanced accuracy \cite{brodersen2010balanced} as our evaluation metric to obtain a better estimate of the model performance. We observe that balanced accuracy scores highly correlate with equal error rate ($EER$) and average cost ($C_{avg}$), which we do not report for the sake of conciseness. 


\subsection{Cross-Domain Evaluation}

Table \ref{table:crossdomain} presents the results of the cross-domain evaluation on both datasets without adaptation. Even though our LID models are not heavily regularized, the in-domain performance is always above 95\%, while MFSC and MFCC features yield a comparable performance. On the other hand, out-of-domain  (OOD) evaluation shows a considerable drop in accuracy in each cross-domain setting. It is interesting to observe that the drop in accuracy is more pronounced for MFCC features, and MFSCs seem to be more robust under domain shift. The impact of domain shift is more pronounced in the GRS $\rightarrow$ RBS direction.

\begin{table}[t]
\caption{Cross-domain evaluation of LID models in acc. (\%).}
\label{table:crossdomain}
\centering
\begin{tabular}{@{}ccccc@{}}
\toprule
\multicolumn{2}{c}{}             & \multicolumn{3}{c}{\textbf{Evaluation}} \\ \cmidrule(l){3-5} 
\textbf{Dataset}     & \textbf{} & In-domain     & OOD       & $\Delta$    \\ \midrule

\multirow{2}{*}{GRS} & MFSCs     & 95.27         & \textbf{43.27}     & -54.55      \\
& MFCCs     & 95.81         & 37.80     & -60.54      \\ \midrule

\multirow{2}{*}{RBS} & MFSCs     & 95.34         & \textbf{54.01}     & -43.35      \\
                     & MFCCs     & 95.00         & 50.97     & -46.35      \\ \bottomrule 
\end{tabular}
\end{table}
 
\subsection{Adaptation Results}
In our adaptation experiments, we investigate two transfer tasks; GRS $\rightarrow$ RBS and RBS $\rightarrow$ GRS. The results are shown in Table \ref{table:adaptation}. The adapted models consistently improve the accuracy compared to the source-only non-adapted baseline with both features and in both directions. Our \textsc{DA-LID II} model yields the best results, which suggests that the domain discrepancy is present not only in the convolutional layers, but also in the fully-connected layers that are more distant from the input. We present and discuss the results for both directions. 

\begin{table}[!b]
 \caption{OOD performance of adapted models in acc. (\%).}
 \label{table:adaptation}
 \centering
\begin{tabular}{@{}ccccc@{}}
\toprule
                                     &                   & \multicolumn{3}{c}{\textbf{Adaptation Method}} \\ \cmidrule(l){3-5} 
\textbf{Direction}                   & & None     & \textsc{DA-LID I}         & \textsc{DA-LID II}     \\ \midrule
\multirow{2}{*}{\footnotesize{GRS} $\rightarrow$ \footnotesize{RBS}} & \footnotesize{MFSCs}             & 43.27    & 47.22  \footnotesize{(+09.1)}    & \textbf{50.56} \footnotesize{(+16.6)}   \\
                                     & \footnotesize{MFCCs}             & 37.80    & 41.77 \footnotesize{(+12.6)}     & 44.55 \footnotesize{(+17.9)}   \\ \midrule 
\multirow{2}{*}{\footnotesize{RBS} $\rightarrow$ \footnotesize{GRS}} & \footnotesize{MFSCs}             & 54.01    & 72.12 \footnotesize{(+33.5)}     & 86.99 \footnotesize{(+61.1)}   \\
                                     & \footnotesize{MFCCs}             & 50.97    & 66.50 \footnotesize{(+30.5)}    & \textbf{90.56} \footnotesize{(+77.7)}   \\ \bottomrule
\end{tabular}
\end{table}

\vspace{0.1cm} 

\noindent
\textbf{RBS $\rightarrow$ GRS} \hspace{0.2cm} Both adapted models yield significant improvements over non-adapted models. The MFCC-based \textsc{DA-LID II} boosts OOD accuracy from 50.94\% to 90.56\% with a relative accuracy gain of 77.7\%. 

\noindent
\textbf{GRS $\rightarrow$ RBS} \hspace{0.2cm} Even though adapted models improve over the baseline, the improvements in this direction are less impressive than what is observed in the  RBS $\rightarrow$ GRS direction. Our MFSC-based \textsc{DA-LID II} model performs best and improves the accuracy by 16.6\% compared to the baseline.
\subsection{Discussion}
The performance gap between the two directions in our experiments seemed surprising at the beginning. In retrospective, this should not be surprising as the two directions are not equally challenging. The RBS dataset is more diverse in terms of the number of unique speakers and background noise. An LID model trained on the RBS dataset has learned to extract language ID features from noisy speech signals, thus it is expected to be more generic and perform well on clean speech signals even under domain shift. This finding is consistent with what has been reported in the domain adaptation literature on how source domain diversity affects adaptability of the model to new domains \cite{ganin2015unsupervised}. On the other hand, if the model has not been exposed to noisy speech signals during training, it is unlikely to perform well on noisy signals even if the representation discrepancy has been minimized, which is the case in the GRS $\rightarrow$ RBS direction. This suggests that alternative adversarial training procedures that add noise to the input representation could be explored to encourage the model to transform the noisy input signals into noise-robust representations. Moreover, our experiments show that MFCCs are more sensitive to input variations due to domain shift, thus MFCC-based models in both directions tend to benefit more from adaptation in terms of relative accuracy gain compared to their MFSC-based counterparts, with only one exception case. 


\section{Adaptive Model Analysis}
In this section, we seek to understand why unsupervised adaptation with adversarial training improves OOD performance. We analyze the results of the RBS $\rightarrow$ GRS transfer task to get insights into the factors that lead to the significant improvement. 
\subsection{Fine-grained Performance Analysis}
\begin{figure}[t]
  \centering
  \includegraphics[width=\linewidth]{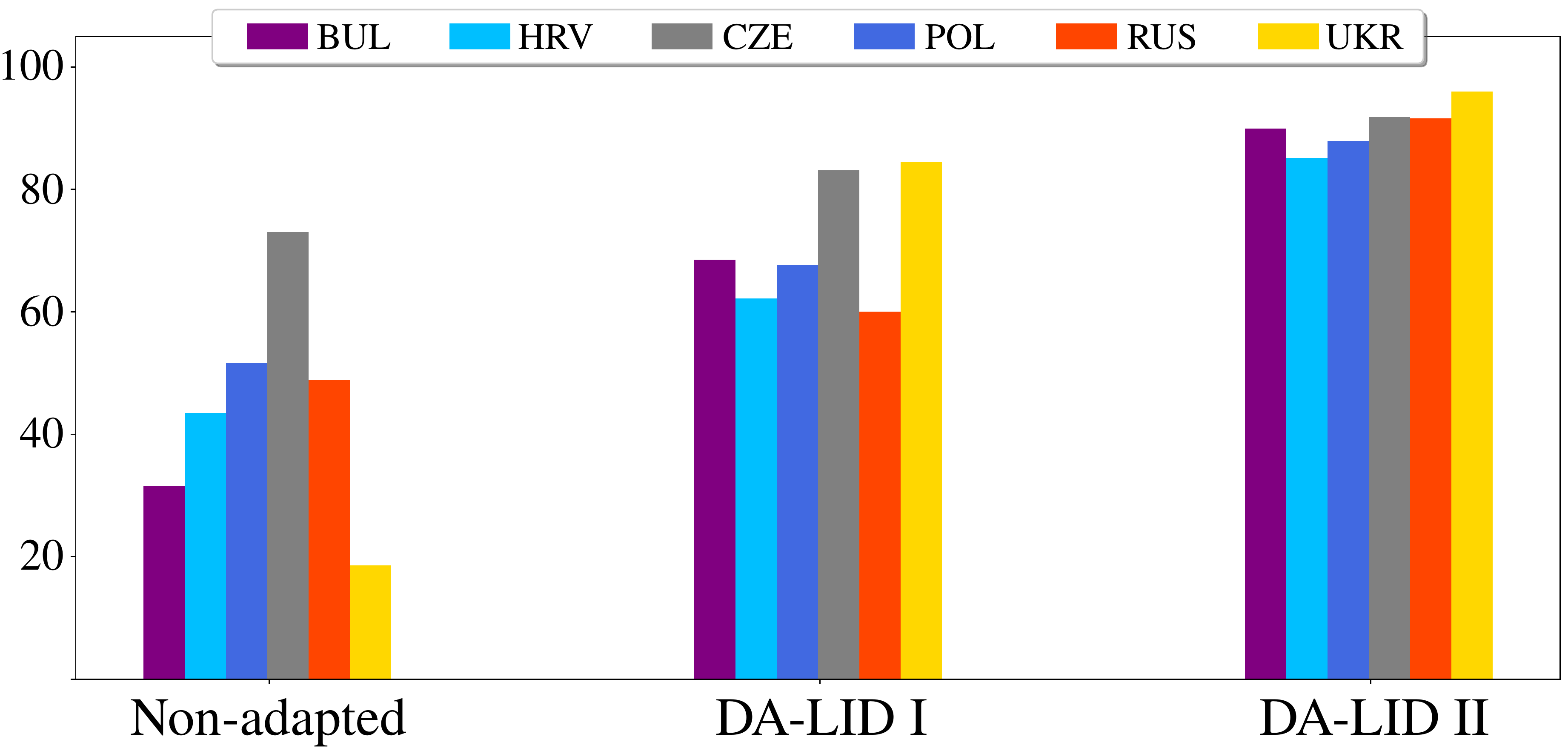}
  \caption{OOD $F_1$ score (\%) per language of our MFCC-based models in the RBS $\rightarrow$ GRS direction.}
  \label{fig:finegraine}
\end{figure}
Fig. \ref{fig:finegraine} shows the performance per language measured by $F_1$ score. In the non-adapted case, we observe a much higher variance between languages compared to the adapted models. For example, while the non-adapted model achieves up to 70\% $F_1$ on Czech, it drops to 18.6\% on Ukrainian, which is slightly better than the chance-level $F_1$  (16.7\%). We inspected the performance on Ukrainian in the other direction and found that the $F_1$ is even worse than chance-level. We hypothesized that the acoustic conditions of the Czech recordings in the two domains are similar, while the discrepancy is maximal in the case of Ukrainian. To validate this hypothesis, we manually inspected several Ukrainian utterances from the GRS dataset. We found that most utterances are characterized by unnatural pauses and hesitations that distort the speech signal and are uniformly distributed across Ukrainian training and evaluation speakers in the GRS dataset. This effect adds to the discrepancy due to domain shift since RBS utterances are more naturally flowing speech than the read speech from the GRS dataset, despite the occasional background noise. In particular, this effect creates abnormal patterns that hinder non-adapted LID performance in two ways: (1) if these patterns are not uniformly distributed across languages and observed during training, the network exploits them as shortcuts to discriminate between languages, and (2) if these patterns are encountered during OOD inference, the distorted signal causes a failure because the model has not been exposed to such patterns during training. Both cases lead to poor OOD generalization when training on a single-domain dataset. However, since these patterns are only present in one dataset, they are good predictors of the domain. Therefore, adversarial training with domain confusion prevents the models from exploiting such dataset-specific artifacts, which consistently yields a better OOD generalization. The advantage of adversarial training is demonstrated in Fig. \ref{fig:finegraine}. Our adapted model boosts the $F_1$ score on Ukrainian from 18.6\% to 96.0\%, which is surprisingly the highest in this direction.
\subsection{Visualizing the Representations}
\begin{figure}[t]
  \centering
  \includegraphics[width=\linewidth]{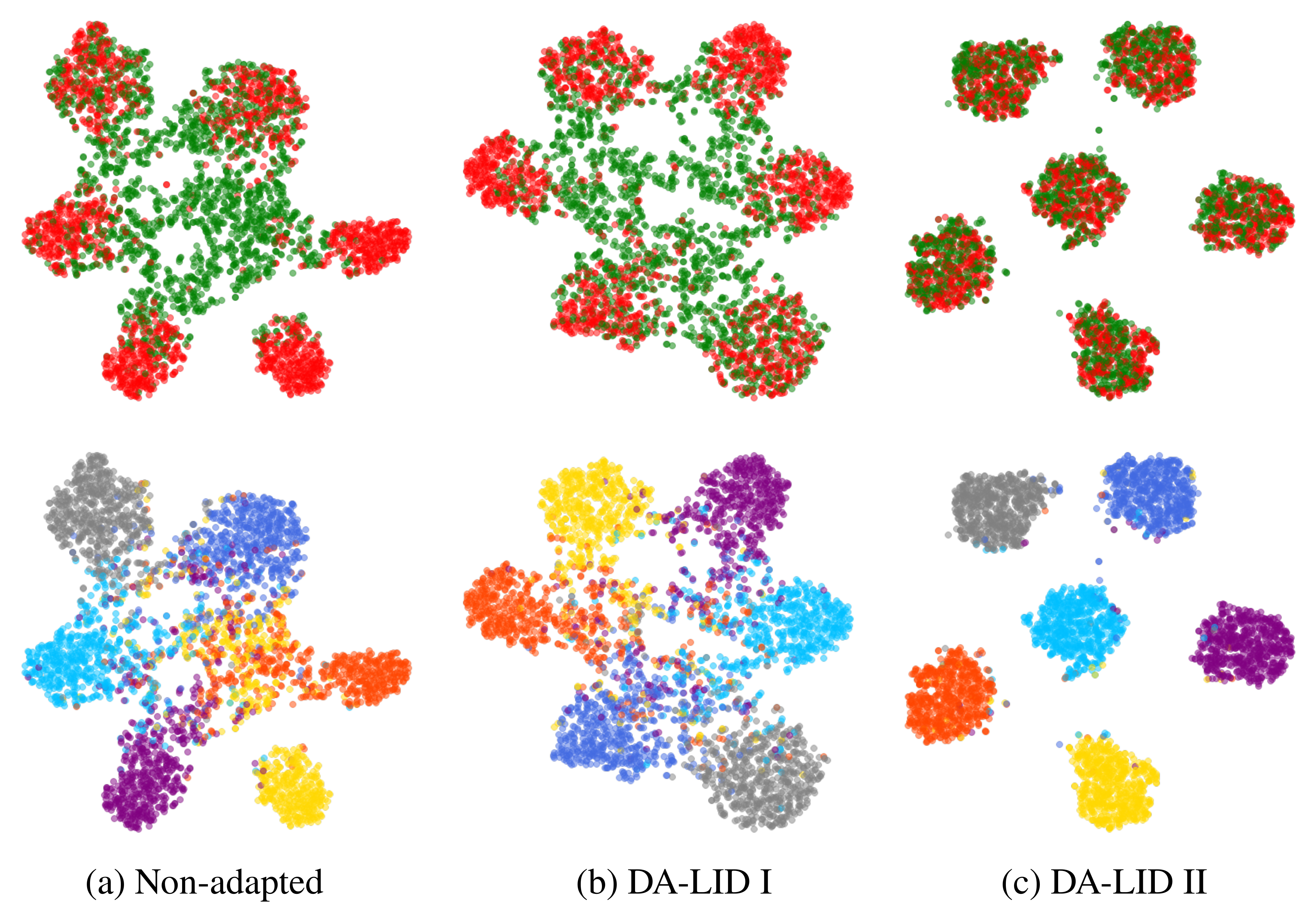}
  \caption{t-SNE visualization: (Top) data points are marked by domain, red points correspond to source samples while green points corresponds to target samples, and (Bottom) data points are marked by language.}
  \label{fig:tsneviz}
\end{figure}

In Fig. \ref{fig:tsneviz}, we visualize the representations using the t-SNE algorithm \cite{maaten2008visualizing}. We sample a set of 1800 data points from each domain and obtain the representations from the last hidden layer of the MFCC-based LID models: (a) source-only non-adapted LID, (b) \textsc{DA-LID I}, and (c) \textsc{DA-LID II}. Fig. 3 shows how adaptation aligns the distributions of the extracted representations from the two domains. 


\section{Conclusions}
We have investigated the problem of spoken language identification for closely-related languages in a cross-domain setting, using deep convolutional neural networks as discriminative models. While our experiments have confirmed that they perform very well within-domain, our cross-domain evaluation has revealed that neural models poorly generalize to a novel dataset with acoustic conditions that differ from those that have been observed during training. To improve the robustness of our models against domain mismatch, we have applied unsupervised domain adaptation with gradient reversal and shown that our adaptive models generalize better across domains. Our analysis has shown that adversarial training prevents the model from exploiting dataset-specific artifacts, thus leading to better out-of-domain generalization. We have identified the diversity of the speech samples in the source domain as the major factor that affects the adaptability of the model to a new domain. Given a diverse source dataset, our adaptive models achieved relative accuracy improvements of up to 77.7\%.

\section{Acknowledgements}

We would like to thank the anonymous reviewers for their insightful suggestions and comments. We extend our gratitude to Marius Mosbach for his valuable feedback and fruitful discussions on this research. This research is Funded by the Deutsche Forschungsgemeinschaft (DFG, German Research Foundation), Project ID 232722074, SFB 1102. 

\bibliographystyle{IEEEtran}
\bibliography{paper.bib}

\end{document}